\documentclass[journal=jacsat,manuscript=communication,layout=twocolumnm]{achemso}
\usepackage{graphicx}
\usepackage{amsmath,amssymb}
\usepackage{mathrsfs}
\usepackage{color}
\usepackage{afterpage}
\usepackage[version=3]{mhchem}
\usepackage[normal]{subfigure}

\title{Strengthening gold-gold bonds \\ by complexing gold clusters with noble gases}
\author{Luca M. Ghiringhelli and Sergey V. Levchenko}
\affiliation{ Fritz-Haber-Institut der Max-Planck-Gesellschaft,
Faradayweg 4-6, D-14195 Berlin, Germany}
\email{ghiringhelli@fhi-berlin.mpg.de, levchenko@fhi-berlin.mpg.de}
\begin{document}

{\em {\bf Abstract.} 
We report an unexpectedly strong and complex chemical bonding of rare-gas atoms to neutral gold clusters. The bonding features are consistently reproduced at different levels of approximation within density-functional theory and beyond: from GGA, through hybrid and double-hybrid functionals, up to renormalized second-order perturbation theory. The main finding is that the adsorption of Ar, Kr, and Xe reduces electron-electron repulsion within gold dimer, causing strengthening of the Au-Au bond. Differently from the dimer, the rare-gas adsorption effects on the gold trimer's geometry and vibrational frequencies are mainly due to electron occupation of the trimer's lowest unoccupied molecular orbital. For the trimer, the theoretical results are also consistent with far-infrared multiple photon dissociation experiments.}\\

Binding of rare-gas (RG) atoms to molecules and metal clusters has been studied intensively in the past years~\cite{Bartlett1962,Holloway1999,Pyykko1995,Graham2000,Khriachtchev2000,Breckenridge2008,Belpassi2008,Gruene2008}.
Due to their very stable closed-shell electronic configuration, when interacting with neutral species and clusters, RG atoms are expected to interact via dispersion forces and polarization by multipole moments of the clusters.

However, a closer look to the binding of RG atoms to Au clusters reveals a much more complex nature of the interaction.~\cite{Pyykko1995,Breckenridge2008,Belpassi2008} In most of the previous studies, electrostatic effects are expected to play a major role in the interaction between the RG atom(s) and Au, since the Au atom is either charged or is a part of a polar molecule (with a large dipole moment). Reports on theoretical analysis of the binding of RG atoms to bare neutral Au$_M$ clusters are very scarce~\cite{jpca}. 
We find that the present understanding of RG-Au$_M$ interaction is not conclusive due to the sensitivity of RG-Au$_M$ electronic structure to the level of theory. Therefore, in this work we analyze bonding between RG atoms (Ne, Ar, Kr, and Xe) and the smallest Au$_M$ clusters ($M$ = 2, 3) using a variety of different theoretical approaches. In fact, we believe that interaction of this type has been overlooked so far, at least for this class of system. Our analysis explains several puzzling features of the observed spectra obtained with far-IR resonance-enhanced multiple photon dissociation (FIR-MPD) spectroscopy, which are reported in detail in Ref.~\cite{long}.

\begin{figure}[t!]
\centering
\includegraphics[width=0.85\columnwidth,clip]{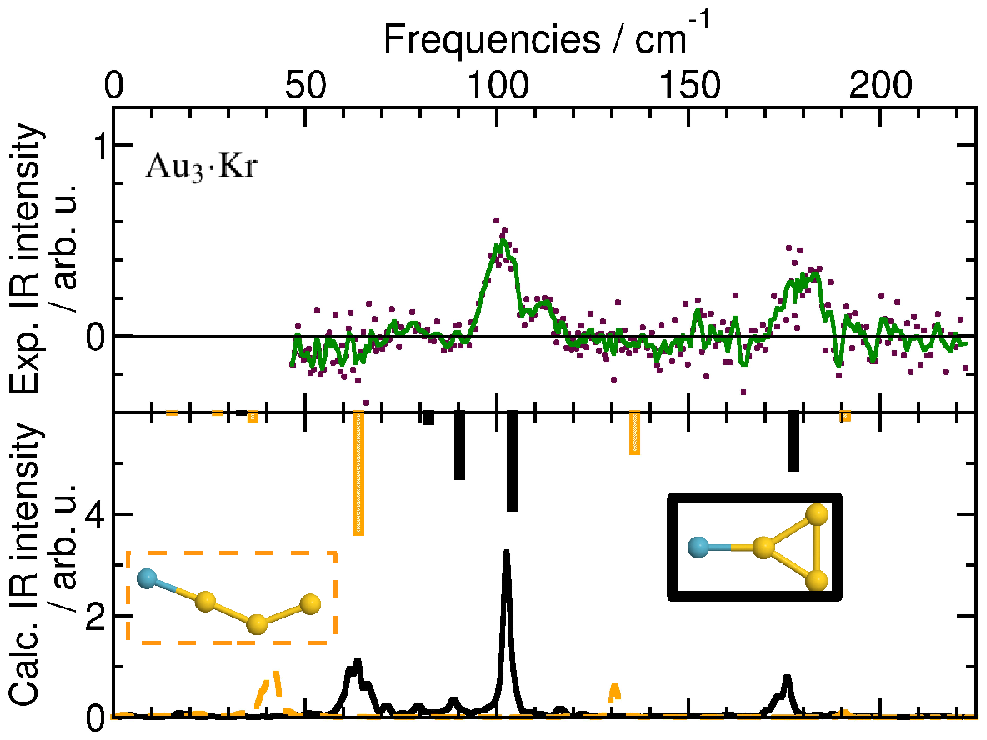}\\
\includegraphics[width=0.85\columnwidth,clip]{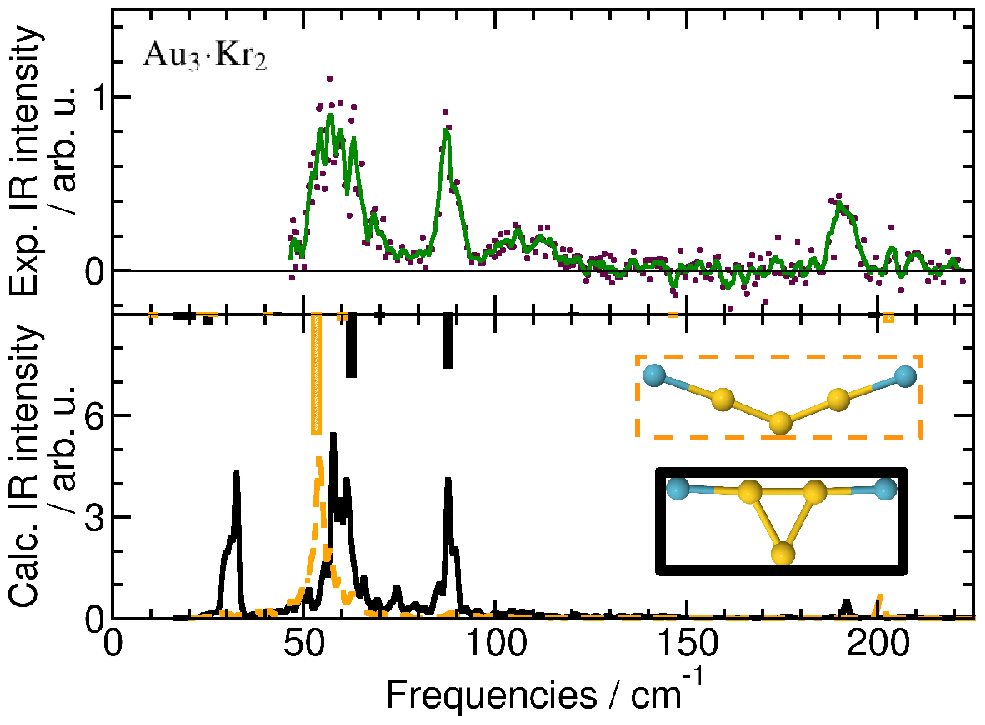}
\caption{FIR-MPD \cite{long} (upper panels) and theoretical IR spectra  \cite{long} (lower panels) at $T=100$~K. Theoretical harmonic spectra are also shown (as upside-down bars in the lower panels) for the two isomers of Au$_3$Kr (top) and Au$_3$Kr$_2$ (bottom). The yellow (dashed) lines refer to the obtuse-angled isomer, while the black lines refer to the acute-angled one.}
\label{F:Au3Kr}
\end{figure}

In Fig. \ref{F:Au3Kr} we show, in the upper half of each panel, the FIR-MPD spectra of Au$_3$ complexed with one or two Kr atoms, at $T$=100~K. The first striking aspect of the measured spectra is that the adsorption of the second Kr changes the spectrum significantly. This fact suggests that the interaction of Kr with Au$_3$ cannot be treated as a perturbation. 

To explain these findings, we have calculated the geometry, electronic structure, and vibrational frequencies of RG-Au complexes. The calculations were performed with  FHI-aims~\cite{Blum2009} program package for an accurate all-electron description based on numeric atom-centered basis functions. 
The IR spectra at finite temperature were calculated by performing Born-Oppenheimer molecular dynamics simulations in the \emph{canonical} ensemble and extracting from the trajectories the Fourier transform of the dipole-dipole autocorrelation function \cite{long}. 
The computational details are given in \cite{long} and SI. 

We first address the simplest system where we observe an unusually strong bond between Kr and a neutral, non-polar species: Kr adsorbed on Au$_2$ (note that adsorption of Kr on a single neutral gold atom leads to a weakly bound van-der-Waals, vdW, complex). 
Au$_2$Kr's geometry is linear and the Kr-Au equilibrium distance is found to be around 2.7~\AA, i.e., in the range of covalent bonding for gold atoms. Furthermore, upon adsorption of one Kr atom, the Au-Au distance is shortened by 0.3\% and the IR spectrum becomes active (see Fig. 1 and Table I in SI), with a Au-Au stretching-mode frequency 5\% blue-shifted with respect to the (IR-inactive) corresponding mode of the isolated dimer. We interpret these results as an indication of the strengthening of the Au-Au bond. To our surprise, both Mulliken and Hirshfeld analysis of the charge transfer (with both PBE and PBE0 functionals) agree in assigning a net charge transfer from Kr to the dimer by 0.1~$e^-$. Such transfer of negative charge from Kr to Au was also predicted in Ref.~\cite{jpca} on the basis of natural bond orbital analysis. This charge transfer is inconsistent with the strengthening of the Au-Au bond; in fact, the Au-Au bond in the Au$_2^-$ is weaker (i.e., longer and with softer stretching frequency, see SI) than in Au$_2$. We find similar contradiction for Au$_3$Kr$_N$ complexes. Interestingly, the highest-frequency experimental peak (around 170~cm$^{-1}$) in Fig. \ref{F:Au3Kr} blue-shifts upon second Kr adsorption.

Geometries and binding energies obtained via PBE+vdW (i.e., the PBE~\cite{PBE} functional, corrected for long-range vdW interactions via the Tkatchenko-Scheffler (TS) scheme\cite{Tkatchenko2009}) are compared for Au$_2$ and Au$_2$Kr to PBE0+vdW and B3LYP hybrid functionals, M06-2X functional~\cite{M06-2X}, XYG3 doubly hybrid functional~\cite{xyg3}, renormalized second-order perturbation theory (rPT2)~\cite{xinguorpa}, MP2, and CCSD(T). 
The molecular orbitals of Au$_M$Kr$_N$ clusters were also calculated with many-body perturbation theory methods $G_0W_0$@PBE0 and self-consistent $GW$ (sc-$GW$), as implemented in FHI-aims~\cite{xinguoRI,Fabio}. CCSD(T) calculations were performed using Gaussian03 code~\cite{g03} with the aug-cc-pVTZ-PP basis set \cite{basisccsd2} 
and small-core relativistic pseudopotentials \cite{basisccsd1}.

%
The calculated Au$_2$Kr binding energies and geometries are summarized in Table \ref{T:summary}. Compared to CCSD(T), we find that PBE+vdW performs reasonably well (3 \% error in binding energies, and 1\% in error bond lengths) and the XYG3 functional performs remarkably well.

\begin{table}[h!]
\centering \footnotesize
\begin{tabular*}{\columnwidth} { l | c | c | c | c | c |}
 & \multicolumn{2}{|c|}{Au$_2$}  & \multicolumn{3}{|c|}{Au$_2$Kr} \\
  \hline
 & $E_b$ \rule{0pt}{2.ex} & $d_\textrm{Au-Au}$ & $d_\textrm{Au-Au}$ &  $d_\textrm{Au-Kr}$ & $\Delta E_b$ \\
 & [eV] & [\AA] & [\AA] & [\AA] &  [eV] \\
\hline  
PBE+vdW\cite{long} \rule{0pt}{2.5ex} & 2.354 & 2.509 & 2.503 & 2.728 & 0.222 \\
PBE0+vdW\cite{long} & 2.050 & 2.520 & 2.514 & 2.773 & 0.161 \\
B3LYP & 2.055 & 2.530 & 2.528 & 2.878 & 0.070 \\
M06-2X & 1.432 & 2.538 & 2.528 & 3.030 & 0.130 \\
XYG3\cite{long} & 2.296 & 2.486 & 2.480 & 2.740 & 0.215 \\
rPT2@PBE0\cite{long} & 2.202 & 2.500 & 2.496 & 2.785 & 0.208 \\
MP2\cite{long} & 2.445 & 2.429 & 2.421 & 2.620 & 0.379 \\
CCSD(T)\cite{long} & 2.292 & 2.484 & 2.477 & 2.685 & 0.320 \\
Exp. \cite{James,Davis} & 2.30  $\pm$ 0.1 & 2.470 & - & - & - \\
\hline
\end{tabular*}
\caption{Calculated properties of Au$_2$ and Au$_2$Kr. The structure is relaxed at each level of theory, for the collinear geometry.
The distance $d_\textrm{Au-Kr}$ is the distance between Kr and the closest Au atom. 
The binding energy of Au$_2$ is $ E_b(\mathrm{Au}_2) = E(\mathrm{Au}_2) - 2 E(\mathrm{Au})$.
The adsorption energy of the Kr atom(s) onto the Au$_2$ is $\Delta E_b( \mathrm{Au}_2\mathrm{Kr} ) = E(\mathrm{Au}_2\mathrm{Kr}) - E(\mathrm{Au}_2) - E(\mathrm{Kr})$, where $E(.)$ is the total energy of the \textit{relaxed} system. For all methods except PBE+vdW the energies are counterpoise corrected for the basis set superposition error.}
\label{T:summary}
\end{table}

{\em All} the employed methods agree in predicting {\em shortening} of the Au-Au bond upon Kr attachment. 
Consistently, the harmonic vibrational frequency of the Au-Au stretch is increased (see Table I in SI). The binding energy of Kr to Au$_2$ is in the range 0.2-0.3~eV, which is much stronger than typical vdW binding ($0.02$ eV per Au-Kr pair, for this system, according to the TS scheme)

The adsorption of other RG species was also considered. We find that Ne, Ar, Kr, and Xe bind by 0.02, 0.11, 0.22, 0.43 eV, respectively (with PBE0+vdW). In all cases the adsorption leads to a small but systematic shortening of the Au-Au bond and increase of the stretching frequency. 

\begin{figure}[h!]
\centering
\includegraphics[width=0.98\columnwidth,clip]{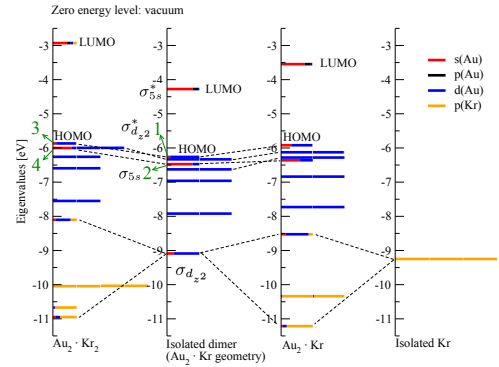}
\caption{Molecular-orbital picture based on atom- and angular-momentum-projected (PBE) electronic density of states, calculated for Kr bonding to Au$_2$. The dashed lines link similar orbitals in different molecular arrangements. 
The labels 1 to 4 denote orbitals used to calculate the change in the partial electron density upon addition of two Kr to Au$_2$, shown in Fig. \ref{F:deltarho}. 
}
\label{F:MOAu2}
\end{figure}

\begin{figure}[t!]
\centering
\includegraphics[width=0.85\columnwidth,clip]{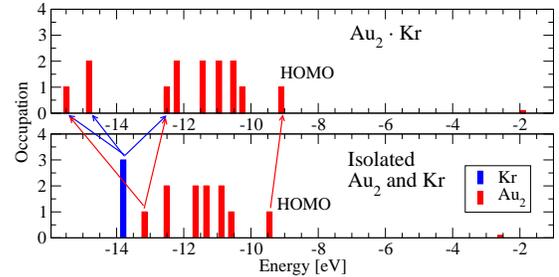}
\caption{Evolution of self-consistent $GW$ spectra from isolated Au$_2$ and Kr to relaxed Au$_2$Kr.}
\label{F:scGW}
\end{figure}

To explain the above results, we analyze the changes in Kohn-Sham molecular orbitals upon Kr binding to Au$_2$. As can be seen from Fig. \ref{F:MOAu2}, the 4$p$ orbitals of Kr atom are close in energy to the $\sigma_{d_{z^2}}$ orbital of Au$_2$, i.e., the linear combination of Au's $d_{z^2}$ orbitals. However, the overlap of these orbitals alone cannot lead to a covalent bonding, because both orbitals are fully occupied. This picture is not an artifact of a particular functional: in fact it is confirmed at the PBE0, $G_0W_0$@PBE0 (Figs. 4 and 5 in SI), and sc-$GW$ (Fig. \ref{F:scGW}) levels.
Following the RG series, we note (see Fig. 2 in SI) that the $p$ valence orbital of Ar, Kr, and Xe, has an energy (with respect to the vacuum level) that is close to the energy of the $\sigma_{d_{z^2}}$ orbital of the dimer.  This is not the case for Ne and, as explained below, this is the reason for the much weaker (vdW only) bonding of Ne to Au$_2$.

Upon Kr adsorption, we find a reduction in the contribution of Au $5s$ orbitals into the $\sigma_{d_{z^2}}$ orbital of Au$_2$(see Fig. \ref{F:MOAu2}). By analyzing differences of orbital densities, we conclude that this reduction corresponds to a polarization of the $\sigma_{d_{z^2}}$ orbital towards Kr. The polarization results in accumulation of the electronic charge between Kr and Au$_2$. The polarisation does not occur for Ne because the mixing of Ne $2p$ and Au2 $\sigma_{d_{z^2}}$ is inhibited by the large difference in energies between these two orbitals.   

In order to quantify this accumulation, we integrated the total electron density difference between Au$_2$Kr and Au$_2$ (at the geometry of Au$_2$Kr), within two planes passing through Au and Kr atoms and perpendicular to the molecular axis. With PBE, the integrated accumulation is 0.012$e^-$. 

\begin{figure}[b!]
\centering
\includegraphics[width=0.95\columnwidth,clip]{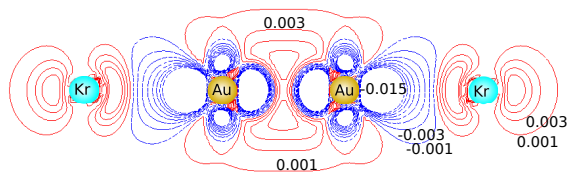}
\caption{Contour plot of the linear combination of the density of the molecular orbitals (MO) of Au$_2$Kr$_2$ and Au$_2$, as labeled in Fig. \ref{F:MOAu2}: (MO$_3+$MO$_4$)$-$(MO$_1+$MO$_2$). The isolevels are equally spaced by $0.002 e^-/$\AA$^3$.}
\label{F:deltarho}
\end{figure}

At the same time, the contribution of the $\sigma ^*_{5s}$ antibonding orbital to the $\sigma^*_{d_{z^2}}$ antibonding orbital increases. This is particularly clear for the linear Au$_2$Kr$_2$ complex, that has the same symmetry (D$_{\infty h}$) as Au$_2$. The adsorption energy of two Kr atoms, $\Delta E_b$, is 0.47~eV, slightly less than twice the adsorption energy of one Kr atom. The Au-Au bond is further strengthened: its length is decreased, and the Au-Au stretching frequency is increased (see Table I in SI). Fig. \ref{F:deltarho} shows that the increased mixing of the occupied and unoccupied antibonding orbitals of Au$_2$ results in the partial removal of the $\sigma^*_{d_{z^2}}$ electrons and placing them to a more diffuse orbital. In other words, the electron-electron repulsion is reduced upon Kr attachment, allowing for more electron density to accumulate between the Au atoms. Overall, this redistribution of orbital electron densities leads to an increase of the total density between the Au atoms. Indeed, using the same method as for the Au-Kr bond, we find an accumulation of 0.030$e^-$ between Au atoms, consistent with the bond strengthening. We attribute the apparent charge transfer from Kr to Au$_2$, as given by both Mulliken and Hirshfeld analysis, to the diffuse nature of the Au$_2$ orbitals, resulting in a misleading assignment of charges to atoms.

We now focus on Au$_3$, that
has two isomers\cite{long}, both isosceles triangles, one obtuse-angled ({\em oa}), with a bond angle of about 140$^\circ$ and the other acute-angled ({\em aa}, bond angle of $\sim 66^\circ$). With PBE+vdW, the {\em oa} isomer is more stable by 0.12 eV. 
By adsorbing one or two Kr atoms to these two isomers, the two resulting structures have approximately the same total energy ({\em aa} isomer less stable by 0.03~eV, with PBE+vdW). 
The final structures of Au$_3$Kr and Au$_3$Kr$_2$ are shown in the insets of Fig. \ref{F:Au3Kr}.
The comparison of theoretical and experimental spectra of Au$_3$ complexed with one and two Kr atoms has been already presented in Ref. \cite{long}.
Here, we focus on the explanation of the blue-shift of the higher frequency peak, related to the {\em aa} isomer, when going from one to two adsorbed Kr (see Fig. \ref{F:Au3Kr}.

Attachment of one Kr to the {\em aa} isomer shortens the opposite Au-Au bond by 0.08 \AA, while adsorption of two Kr atoms shortens the adjacent Au-Au bond by 0.32 \AA. The corresponding adsorption energies are 0.23 and 0.34 eV.
In order to explain the calculated binding energies, the changes in geometry, and the blue-shift of the highest-frequency mode upon Kr attachment to Au$_3$, we perform the analysis similar to the Au$_2$Kr case. The highest-occupied molecular orbital (HOMO) of Au$_3$ is singly occupied. Its character is bonding for the two short bonds and antibonding for the long bond (opposite to the $66^\circ$ angle). The lowest unoccupied molecular orbital (LUMO) of (relaxed) Au$_3$ is antibonding for the short bonds, and bonding for the other bond. In this case, the 4$p$ orbital of single added Kr mixes with the LUMO of Au$_3$, leading to the increased LUMO-$d$-manifold gap, decreased Au-Au distance for the long bond, and increased Au-Au distances for the equivalent short bonds (see Fig. 7 in SI). Kr 4$p$ orbital does not mix with HOMO of Au$_3$ by symmetry. Thus, contrary to Au$_2$ complexes, the bonding can be described as partial donation of a Kr electron pair to the Au$_3$ cluster.
Addition of the second Kr atom leads to swapping of the LUMO and HOMO orbitals (so that the former LUMO becomes singly occupied), and a large reduction ($-11$\%) of the bond distance of the longest bond in Au$_3$ (see Table I and Fig. 6 in the Suppl. Material). The vibrational analysis shows that it is the strengthening of this bond that leads to the blue shift of the observed high-frequency peak.

In conclusion, we find that RG atoms bind covalently to small neutral Au clusters. For closed-shell clusters (Au$_2$ in particular), rather exotic features of this binding become prominent: it involves a redistribution of electrons among the $d$- and $s$-derived antibonding orbitals, leading to the strengthening of the Au-Au bonds. In the open-shell complexes (such as Au$_3$Kr), however, mainly the partial donation of electrons from Kr to the unoccupied orbitals of the Au cluster determines the bonding. Beside fully explaining fine details in the FIR-MPD of the gold trimer, our findings suggest that weakly adsorbing species can fine-tune the electronic properties of very small clusters, which may be of interest for the use of gold clusters as catalysts.

\providecommand*{\mcitethebibliography}{\thebibliography}
\csname @ifundefined\endcsname{endmcitethebibliography}
{\let\endmcitethebibliography\endthebibliography}{}

\end{document}